\documentclass[aps,prl,showpacs,amsmath,amsfonts,superscriptaddress,twocolumn]{revtex4}
\usepackage{graphicx}
\usepackage{dcolumn}
\usepackage{bm}
\usepackage{pslatex}
\begin{document}

\title{Survival of parity effects in superconducting grains at finite temperature}

\author{K.~Van~Houcke}
\author{S.M.A.~Rombouts}
\affiliation{Universiteit Gent - UGent, Vakgroep Subatomaire en
  Stralingsfysica \\
  Proeftuinstraat 86, B-9000 Gent, Belgium}
\author{L.~Pollet}
\affiliation{Theoretische Physik, ETH Z\" urich, CH-8093 Z\" urich, Switzerland}
\date{\today}

\begin{abstract}
We study the thermodynamics of a small, isolated superconducting grain 
using a recently developed quantum Monte Carlo method. 
This method allows us to simulate grains at any finite temperature and with any level
spacing in an exact way.
We focus on the pairing energy, pairing 
gap, condensation energy, heat capacity and spin susceptibility to describe the grain. 
We discuss the interplay between finite size (mesoscopic system), pairing 
correlations and temperature in full detail.
\end{abstract}


\maketitle


The bulk properties of a superconductor are well described by standard BCS
theory. When the system size is reduced however, its mesoscopic behavior is strongly
dictated by the finite electron number. 
For such small systems with a fixed number of
particles, BCS theory is no
longer applicable since the BCS order parameter is identically zero.
Therefore it cannot determine the lower size limit for which the system exhibits superconducting
properties. 
It was suggested by Anderson \cite{Anderson59} that superconductivity would disappear once the
average level spacing $d$ ($\propto 1 / V$, $V$ being
the volume of the system) of the electron spectrum becomes larger than the bulk
superconducting gap $\Delta$.
Due to a series of experiments by Ralph, Black and Tinkham (RBT) \cite{Ralph96}
on
the transport through
a single superconducting nm-scale Al grain, a lot of authors shed new light on
Anderson's suggestion. 
In their experiments, RBT found a
spectroscopic gap
larger than the average level spacing, which goes to zero when applying a
suitable magnetic field. 
The measurements also revealed a peculiar parity effect: 
grains with an even number of electrons have a larger gap in the spectrum than
grains with an odd electron number.  
These observations were regarded as signs of 'superconductivity', in the sense that there is a
pair-correlated ground state.
Properties indicative of strong pairing correlations were only found in grains
with $d  \lesssim  \Delta$.
So Anderson's answer turned out to be uncomplete, since it does not
differentiate between odd and even numbers of electrons.
A  large  number of theoretical studies tried to 
characterize the ground state correlations and 
superconductivity of such small systems in a qualitative way and tried to
predict the critical level spacing at which the superconductivity breaks down.
An extended review can be found in Ref. \cite{vondelft01}.
In this report we study the competition between pairing, finite size and
finite temperature in an exact way, with all quantum correlations taken into account.

To model small superconducting grains, one uses the reduced BCS Hamiltonian
\cite{vondelft96}: 
\begin{equation}
H = \sum^{\Omega}_{\sigma=\pm, j=1} (\epsilon_j - \sigma \mu_B h     )c^{\dag}_{j,\sigma} c_{j,
  \sigma} - \lambda d \sum_{j,j'=1}^{\Omega} B^{\dag}_j B_{j'}, 
\label{eq:BCShamil}
\end{equation}
where $B^{\dag}_j = c^{\dag}_{j,+} c^{\dag}_{j,-}$.
The operator $c^{\dag}_{j,\sigma}$ creates an electron in the single-particle state $|j,\sigma\rangle$.
The quantum number $j$ labels the $\Omega$ single particle levels with energies
$\epsilon_j$, and $\sigma$
labels time reversed states. 
Since the pairing interaction only scatters time-reversed pairs of electrons within
an energy $\omega_D$ of the Fermi level $\epsilon_F$, electrons outside the
cutoff are not taken into account.
 The parameter $\lambda$ is the dimensionless BCS coupling constant and
is related to $\Delta$ and $\omega_D$ via the bulk gap equation $sinh
(1/\lambda) = \omega_D/\Delta$ \cite{Fetter}. 
We take $\lambda=0.224$, close to that of Al \cite{Braun98}. 
The Zeeman term couples an external magnetic field $h$ to the electrons and $\mu_B$
is the Bohr magneton. 
Throughout the paper, we will consider a
half-filled band with fixed width $2 \omega_D$ and
$\Omega = 2 \omega_D/d$ doubly degenerate and uniformly spaced
levels with energies $\epsilon_j = jd$. We will only discuss the case without
magnetic field $h$.

To study the cross-over from the bulk to the few
electron limit, a number of authors originally used a parity-projected
grand canonical (g.c.) BCS approach \cite{vondelft96, smith96, matveev97,
  braun97, Balian98, Balian99}.
The parity effect can be explained with 
this variational technique. 
However,
an artificial sharp
transition to the normal state appears at some critical level spacing and
temperature, which is impossible for a finite system.
Since  the electron number fluctuations are strongly suppressed by charging
effects in the experiments of RBT, it is clear that a canonical formalism is needed to describe the
grains properly.
A number of canonical techniques were used to tackle this problem.
Unfortunately, exact diagonalization techniques (e.g. Lanczos \cite{Mastellone98})
can only handle systems with a very small
number of electrons. 
In order to go to larger model spaces,  particle number projection was combined
with the static path approximation (SPA) plus
random-phase approximation (RPA) treatment \cite{Rossignoli98, Falci02} and with variational wavefunctions  \cite{Braun98}.
Dukelsky and Sierra developed a particle-hole version of the density-matrix
renormalization-group (DMRG) method to study the crossover \cite{Dukelsky99, Dukelsky00}.
All these canonical techniques
reveal the parity effect at low enough temperatures, and make
clear that the abrupt cross-over is just an artefact of the g.c. approach. 
It turned out that small grains with $d \lesssim \Delta$ are indeed
characterized by strong
superconducting pairing correlations. 
As the grain size decreases, quantum fluctuations of the order parameter
start to play a crucial role.
These fluctuations make 
the cross-over completely
smooth without any sign of critical level
spacing.
Only when the grain is not too small 
($d \ll \Delta$) the fluctuations in the order parameter
can be neglected, making the mean field description of
superconductivity appropriate. 
In the
canonical picture, pairing correlations still exist at arbitrary large values
of $d/\Delta$, though in the form of weak fluctuations. 
Qualitative differences between the pairing correlations in the bulk and the few-electron regime make it
still possible to speak of the superconducting regime ($d \ll \Delta$) and the
fluctuation-dominated regime ($d \gtrsim \Delta$) \cite{vondelft01}.

It was only after the appearance of most of these works that one
became aware of the fact that  
the reduced BCS model has an exact solution, worked out decades ago by Richardson in the context
of nuclear physics \cite{Richardson63}. 
In Ref. \cite{Sierra00}, Sierra {\it et al.} 
compare the previously mentioned treatments with the
exact solution.
Using this exact solution to study the finite temperature behavior for a large number of
  many-particle states is difficult due to the exponential scaling of the number of eigenstates that need to be
considered. 
Gladilin {\it et al.}  developed an approximation based on the Richardson
  solution to get finite temperature information \cite{Gladilin}. 
In Ref. \cite{vondelft01} it was already suggested by von Delft and Ralph  that
quantum Monte Carlo (QMC) techniques could be helpful to investigate the BCS
pairing model at finite temperature. Recently we developed a new quantum
Monte Carlo method \cite{Rombouts05, VanHoucke05} that is able to simulate
the BCS model for any fixed number of particles without sign problem. 
The method allows calculating thermodynamic properties in an exact way, up to a controllable statistical
error. Simulations can be performed at any finite temperature and any level
spacing $d/\Delta$ for large system sizes.  
Because
our method  allows a projection on specific symmetries like the total spin
projection, we can calculate the susceptibility and magnetization.

\begin{figure}[h]
\begin{center}
\includegraphics[angle=0, width=7.5cm] {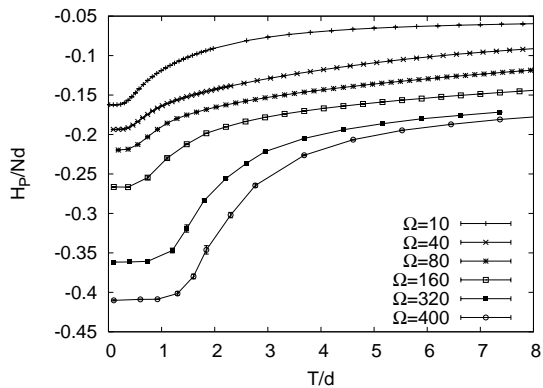}
\caption{  The temperature dependence of the pairing energy per electron for grains with an
  even number $N=\Omega$ of electrons. Simulations were performed for different grain
  sizes $\Omega$. The energy and
  temperature scale is set by the level spacing $d$.  
   \label{fig:ppeven}}
\end{center}
\end{figure}

\begin{figure}[h]
\begin{center}
\includegraphics[angle=0, width=7.5cm] {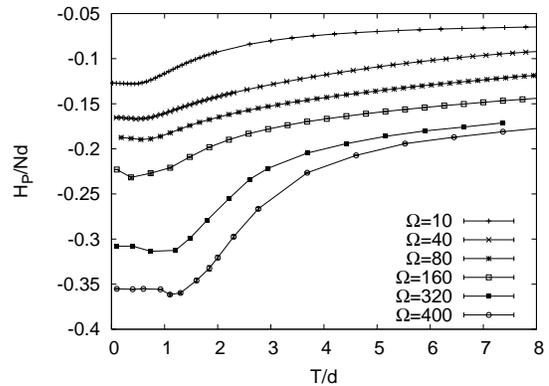}
\caption{ The pairing energy per electron as a function of temperature for an
  odd grain with different sizes. Simulations were performed for grains with $N=\Omega+1$
  electrons. 
   \label{fig:ppodd}}
\end{center}
\end{figure}

\begin{figure}[h]
\begin{center}
\includegraphics[angle=0, width=7.5cm] {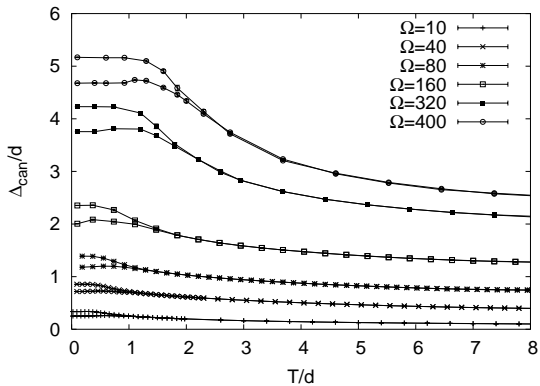}
\caption{ The canonical pairing gap as a function of temperature. 
For each number of levels $\Omega$ the gap
is calculated for an even ($N=\Omega$) and an odd ($N=\Omega+1$) number of electrons. 
Only at low enough temperature one can distinguish between the gap of the even
grain (upper curve) and the odd grain (lower curve) of the same size $\Omega$. 
   \label{fig:gap2}}
\end{center}
\end{figure}

We performed simulations of grains with different sizes ($\Omega$ equal to $10$, $40$, $80$,
$160$, $320$ and $400$).
These half-filled model spaces lead respectively to ratios 
$d/\Delta$ of $8.68$, $2.17$, $1.09$, $0.54$, $0.27$ and $0.22$. 
Figs. \ref{fig:ppeven} and \ref{fig:ppodd} show the thermal averages of the
pairing energy $H_P = - \lambda d \sum_{j,j'=1}^{\Omega} B^{\dag}_j B_{j'}$ per particle as a
function of temperature for even and odd grains. The energy scale is set by the level spacing $d$. 
By comparing both figures, one notices that at low enough temperatures
(typically $T \lesssim d $) the even electron system has more pairing energy than
the odd system. This is due to the single unpaired electron, which blocks the
Fermi level in the odd case. Around
$T \approx d $ a small dip appears in the odd pairing energy.
Qualitatively, this can be explained as follows: due to the thermal energy,
the single unpaired electron is moved one level upward, making the Fermi
level available to pair scattering.   
This is reflected in a
slight decrease of the pairing energy in Fig. \ref{fig:ppodd}.
To measure the real correlation energy due to pairing in the system, 
the 'canonical' pairing gap
\begin{equation}
\Delta_{{\rm can}}^2 = (\lambda d)^2 \sum_{m,n=1}^{\Omega} (\langle B^{\dag}_m B_n \rangle  -
\langle B^{\dag}_m B_n\rangle_{\lambda=0}), 
\end{equation}
was introduced in Eq. (92) of Ref. \cite{vondelft01}. 
The second term subtracts the thermal average of the pairing interaction for
the non-interacting system. When going to the thermodynamic limit,
$\Delta_{{\rm can}}$ becomes equivalent to the BCS bulk gap $\Delta$ \cite{vondelft01}. 
Figure \ref{fig:gap2} shows the even and odd canonical gap
for different system sizes. 
It
follows very clearly that
the temperature scale at which the parity effect appears is set
by the level spacing $d$, and this for all grain sizes.
The crossover temperature 
is given by $T_{cr} = \Delta \ln N_{{\rm eff}}$, with $N_{{\rm eff}}$ the effective number
of states available for excitation ($N_{{\rm eff}} =
\sqrt{8 \pi T \Delta}/d$ in the limit $d\ll \Delta$) \cite{Tuominen92}.
This is in qualitative agreement with Figure \ref{fig:gap2}, where the
crossover temperature decreases as the grain size is reduced.
One should of course keep in mind that the temperature is shown in units of the
level spacing which is considerably smaller for the largest grains.
Figure \ref{fig:gap2} shows that pairing correlations persist even for
ultrasmall grains and that  
a reduction of the grain size leads to a suppression of these
correlations.

\begin{figure}[h]
\begin{center}
\includegraphics[angle=0, width=7.5cm] {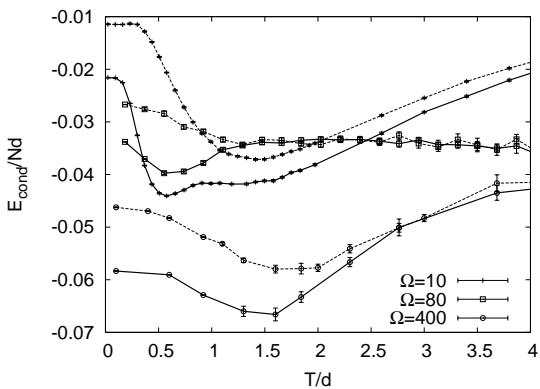}
\caption{ The condensation energy per particle as a function
  of  $T/d$ for system sizes $\Omega=10, 80$ and $400$. Even (odd) grain data points
  are connected by a solid (dashed) line.
   \label{fig:en}}
\end{center}
\end{figure}

The condensation energy $E_{cond} = \langle \psi|H| \psi\rangle - \langle FS|H| FS\rangle$ is the energy difference of the
state $| \psi \rangle $, where all quantum correlations are included,  and the non-correlated
Fermi sea $| FS \rangle $. Figure \ref{fig:en} shows the thermal average of
the condensation energy
per particle for a number of even and odd grains as a function of
temperature. These energy differences were obtained by calculating the thermal averages of the Hamiltonian
over correlated states $|\psi \rangle$ and over the Fermi states
$|FS\rangle$ separately. Below temperatures of the order $d$, the even grains
have a larger condensation energy (in absolute value). Both even and odd
grains have a minimal condensation energy around $T\approx d$. In agreement
with Ref. \cite{vondelft01}, our calculations   give a quasi
intensive condensation energy for  the smallest grains ($d\gg\Delta$), while the condensation energy of grains with
$d \ll \Delta$ increases (in absolute value) inversely proportional to $d$.

\begin{figure}[h]
\begin{center}
\includegraphics[angle=0, width=7.5cm] {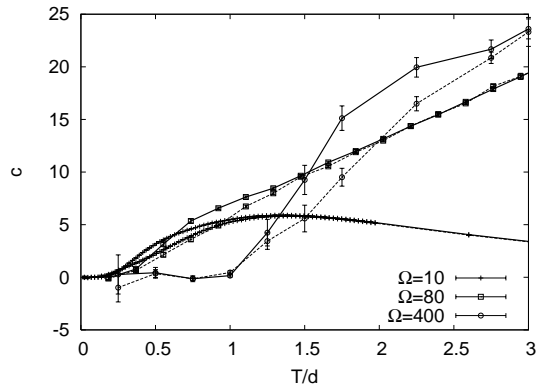}
\caption{ The heat capacity $c = \frac{\partial\langle H \rangle} {\partial T}$ as a function
  of  $T/d$ for system sizes $\Omega=10, 80$ and $400$. 
Even (odd) grain data points
  are connected by a solid (dashed) line.
Around
  temperatures $T \approx 0.5 d$ for $\Omega=10, 80$ and $T \approx d$ for
  $\Omega= 400$ the heat capacity of the even grain (with
  $N=\Omega$ electrons) exceeds the odd ($N=\Omega +1$) specific capacity.
    \label{fig:sh}}
\end{center}
\end{figure}

Figure \ref{fig:sh} shows the heat capacity as a function
of temperature for sizes $\Omega=10, 80$ and $400$. 
Around the crossover temperature where the parity effect becomes visible 
(see Figures \ref{fig:gap2} and \ref{fig:en}), 
a slight parity effect also appears in the heat capacity. Here the even heat
capacity exceeds the odd one. 
At higher temperatures the odd and even results become
indistinguishable again. 
For the $\Omega=10$ grain size, the
finite model space makes the Shottky peak visible when the temperature
becomes of the order of the level spacing. 

\begin{figure}[h]
\begin{center}
\includegraphics[angle=0, width=7.5cm] {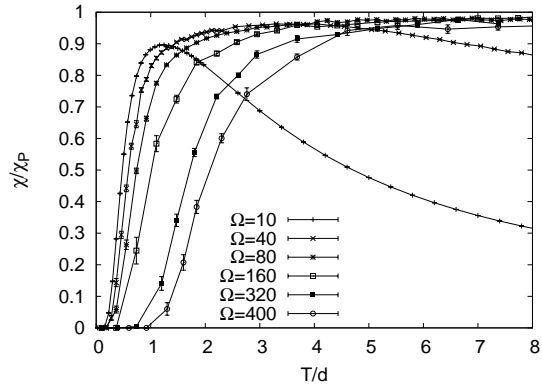}
\caption{ The spin susceptibility normalized to its bulk high temperature
  value $\chi_P$ as a function of $T/d$ for a number of even grains. Each grain contains $N=\Omega$
  electrons, with $\Omega$ the model space size.
   \label{fig:suseven}}
\end{center}
\end{figure}

\begin{figure}[h]
\begin{center}
\includegraphics[angle=0, width=7.5cm] {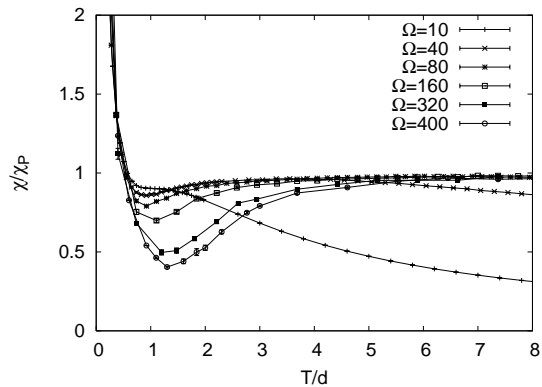}
\caption{ The spin susceptibility as a function of $T/d$ for a number of odd
  grains (containing $N=\Omega+1$ electrons). 
  The grain size is determined by the model space size $\Omega$.
   \label{fig:susodd}}
\end{center}
\end{figure}

The spin susceptibility of a grain is defined by
\begin{eqnarray}
\chi (T) & = &  - \frac{\partial \mathcal{F} (T,h)}{\partial h^2} \bigg|_{h=0}
\nonumber \\
 & = & \frac{1}{T}  (\langle M^2\rangle - \langle M \rangle ^2 ).
\end{eqnarray}
Here $\mathcal{F} = - T \ln Z$ is the free energy of the grain, with $Z$ the canonical partition
function. The susceptibility is proportional to the fluctuation of the 'magnetization'
$M = - \mu_B \sum_{\sigma, n} \sigma 
c^{\dag}_{n,\sigma} c_{n,\sigma}$ 
at finite temperature $T$.
The spin susceptibility of a single isolated grain has been studied by Di Lorenzo
{\it et al.} \cite{Dilorenzo00}.
They found that the pairing correlations affect the
temperature dependence of the spin susceptibility.
In particular, if the number of electrons in the grain is odd, the spin susceptibility 
shows a re-entrant behavior as a function of T for any value of the ratio $d /
\Delta$. 
They show that this behavior persists even in the case of
ultrasmall grains, where the level spacing is much larger than the BCS gap.
 Since this re-entrance behavior is absent in normal metallic grains, they
 suggested that this quantity can be measured and
 used as a unique signature of pairing correlations
in small and ultrasmall grains.
The susceptibility was calculated 
by combining an
analytic analysis in the limiting cases $\Delta \gg d$ and $\Delta \ll d$ 
with
a static path approximation for intermediate values. 
By 
means of exact canonical methods based on Richardson's solution, they also got
exact results at low temperatures. 
With the aid of our QMC method, we
are now able to solve the problem exactly for the whole temperature range.
Figures \ref{fig:suseven} and \ref{fig:susodd} show the temperature dependence
of the spin susceptibility for a number of even and odd grains, respectively.
The susceptibility is normalized to its bulk high temperature value $\chi_P =
2 \mu_B^2/d$. Our results are completely in line with those of Di Lorenzo {\it et al.} \cite{Dilorenzo00}.
At low temperatures the even susceptibility remains exponentially small, while
for an odd grain
the unpaired spin gives rise to an extra paramagnetic
contribution to the spin susceptibility ($\chi \simeq \mu_B^2 / T$). 
The minima in the odd spin susceptibilities coincide with a small increase of
the pairing correlations (see Figs. \ref{fig:ppodd} and \ref{fig:gap2}), with
a mimimal condensation energy (see Fig. \ref{fig:en}) and with a parity effect in
the heat capacity (see Fig. \ref{fig:sh}).
For the smallest odd grain no re-entrant behavior is visible in
Fig. \ref{fig:susodd}. This is an effect of the finite
model space. If the BCS coupling constant is increased a little, a
re-entrance effect appears also in this case.

In conclusion, we solved the BCS pairing problem at finite temperature exactly
via quantum Monte Carlo simulation. 
We studied odd and even grains with a large number of electrons and
arbitrary level spacings.
Our exact results confirm predictions of previous
approximate calculations, showing that the physics
of ultrasmall superconducting grains is well described
by a pairing model with exact particle number projection and that parity
effects are visible in thermodynamic properties.

The number of unpaired electrons in a grain can be increased by an external magnetic field.
Frauendorf {\it et al.} 
showed that at
zero temperature a magnetic field attenuates the pairing, but for a
mesoscopic system in a strong magnetic field the pairing
correlations may come back after heating \cite{Frauendorf03}. 
Such a re-entrance of pairing correlations has also been discussed by Balian
{\it et al.} \cite{Balian99}.
Work on this problem of how an external
field can influence the thermodynamic properties of a single superconducting
grain is in progress.

The authors wish to thank K.Heyde, J. Dukelsky and S. Frauendorf for
interesting suggestions and discussions. We acknowledge the financial support
of the Fund for Scientific Research - Flanders 
(Belgium) and the Swiss National Science Foundation.

\end{document}